\documentstyle[12pt,psfig]{article}

\newcommand{\ra}{\rightarrow}

\newcommand{\be}{\begin{equation}}
\newcommand{\ee}{\end{equation}}

\newcommand{\ba}{\begin{eqnarray}}
\newcommand{\ea}{\end{eqnarray}}

\newcommand{\lda}{\lambda}

\begin{document}


\begin{center}
{\Large{Diffractive $\phi$ Production in a Perturbative QCD Model}
}
\end{center}
\vskip .2in
\begin{center}
J. Alam\footnote{E-mail : alam@nt.phys.s.u-tokyo.ac.jp},
A. Hayashigaki\footnote{E-mail : arata@nt.phys.s.u-tokyo.ac.jp},
K. Suzuki\footnote{E-mail : ksuzuki@nt.phys.s.u-tokyo.ac.jp},
and T. Hatsuda\footnote{E-mail : hatsuda@nt.phys.s.u-tokyo.ac.jp}
\end{center}
\vskip .1in
\begin{center}
{\it Department of Physics, University of Tokyo, Tokyo 113 0033, Japan
     }\\
\end{center}

\parindent=20pt
\vskip 0.1 in
\begin{abstract}

The elastic leptoproduction of $\phi$ measured by the H1 
collaboration at HERA is described by a perturbative QCD model,
based on open $s\bar{s}$ production and parton hadron duality, 
proposed by Martin et al. 
We observe that both the total cross section
and the ratio of the longitudinal and transverse cross sections
are well reproduced with an effective
 strange-quark mass $m_s\,\sim\,320- 380 $ MeV for various
  gluon distribution functions.  
   Possible connection of the effective mass and the 
  momentum dependent dynamical mass associated with
   dynamical breaking of chiral symmetry is also discussed.

\end{abstract}

\vskip 0.2in
\noindent{PACS: 13.85.Ni,14.40.Gx,12.38.Bx,12.39.Jh}
\vskip 0.2in

\section*{I. Introduction}
\begin{figure}
\centerline{\psfig{figure=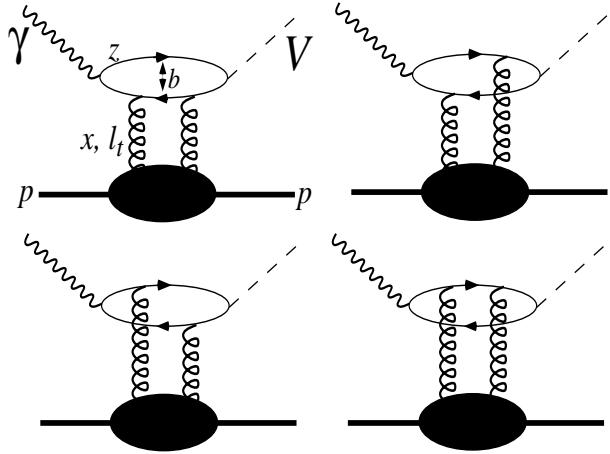,height=6cm,width=8cm}}
\caption{The diffractive vector meson production in $e\,p$
collisions. $b$ is the separation between the quark and the
anti-quark. $x$ and $l_t$ $(x^\prime$ and $-l_t$, not shown in the figure) 
are the Bjorken-$x$ and transverse momentum of the
left (right) gluon respectively.
}
\label{feynman}
\end{figure}

Recently the diffractive photo and leptoproduction of vector mesons
in electron proton collisions ($e\,+\,p\,\ra\,e\,+\,\gamma^\star\,+\,p^\prime 
\,\ra\,e\,+\,V\,+\,p^\prime$) has drawn considerable attention
 from experimental (see {\it e.g.} ~\cite{ha}) 
as well as theoretical side~\cite{ah,ad,bzk,jn,sjb,ir,ks,ph}.  
As the cross section for the diffractive vector
meson production depends (quadratically) on the 
gluon distribution of the proton, it gives a  
unique opportunity to study the low $x$ behavior of the gluons 
inside the proton and to investigate
the transition from perturbative to non-perturbative 
region.  Experimental data 
for vector meson production at HERA~\cite{ha,epj10} 
in the reaction $e\,p\ra V\,e\,p$
are available over a wide range of the virtuality of the photon, $Q^2$,
the {\it cm} energy of $\gamma^{\star}p$ system, $W$, mass of  the 
vector mesons, $m_V$ and the  
four momentum transfer, $t$ in the process. 
The physical picture for the vector meson production
is demonstrated through the diagrams in fig.~\ref{feynman}. 
The virtual
(or real) photon fluctuates to quark anti-quark ($q\bar{q}$) pairs,
which interacts with the target proton via two 
gluons exchange. This interaction changes the transverse momenta of the pair,
which subsequently hadronizes to a vector meson.  It can
be shown~\cite{wm} that the time scale for the interaction of the $q\bar{q}$
with the proton is considerably smaller than the time scale
for the $\gamma^\star\,\ra\,q\bar{q}$ dissociation and vector
meson formation. This leads to the following factorization 
for the amplitude of diffractive vector meson production,
\be
A(\gamma^\star\,p\ra\,V\,p)=\psi_{q\bar{q}}^{\gamma^\star}\otimes
\,A_{q\bar{q}+p}\otimes\,\psi_{q\bar{q}}^V ,
\label{amp} 
\ee
where $\psi_{q\bar{q}}^{\gamma^\star}$ is the wave function
of the virtual photon in $q\bar{q}$, 
$A_{q\bar{q}+p}$ is the
amplitude for the $q\bar{q}$-$p$ interaction and 
$\psi_{q\bar{q}}^V$ is the vector meson wave function.

The process under consideration is  governed by the
scale $K^2\,\sim\,z(1-z)(Q^2+m_V^2)$, indicating that 
for large $Q^2$ and/or $m_V^2$ the perturbative
QCD (pQCD) is applicable to describe the diffractive
vector meson production~\cite{sjb,jcc}.
For the production of heavier mesons ($J/\Psi$ or $\Upsilon$) 
such process is under control by pQCD even for 
$Q^2=0$ (photo-production). A reliable description 
for the heavy vector meson production 
is obtained by using eq.~(\ref{amp}),
which involves vector meson wave functions~\cite{bzk,jn,sjb,lf}. These 
wave functions can be 
obtained by solving Schr\"odinger equation 
with non-relativistic potential model~\cite{eje}.
It has been argued in ~\cite{adm1}
that the main uncertainty to the description of the
light vector mesons ($\rho$) production (particularly 
in the transverse cross section) originates
 from its wave function. To avoid this problem, 
Martin et al.~\cite{adm1} proposed 
a model based on the open $q\bar{q}$ production and 
parton-hadron duality to describe various features of
$\rho$ production in diffractive processes. Subsequently
the same approach has been used to study the 
diffractive $J/\Psi$~\cite{adm2} and $\Upsilon$ 
production~\cite{adm3}. 

Very recently the experimental data for
elastic electroproduction of $\phi$ mesons
at HERA has been made available by the 
H1 collaboration~\cite{h1phi}. In the present article we follow 
Ref.~\cite{adm3} to study the $\phi$ production at HERA energies.
The sensitivity of the results on the strange quark mass is
examined. It is found in our analysis that the experimental
data is described well with an effective strange quark
mass, interpolated between the current and the constituent mass.
In this approach one first calculates the amplitude 
for the open $s\bar{s}$ production, then takes the
projection of the amplitude on the 
 $J^P=1^{-}$ state appropriate for $\phi$
quantum numbers and finally integrating over  
an invariant mass interval such that it 
contains the resonance peak for the vector meson, 
$\phi$, here. 

The paper is organized as follows. In the next
section we discuss the model used in the present
work. Section III is devoted to present the results
and finally in section IV we give summary and conclusions.

\section*{II.  pQCD Model}
The differential cross section 
for the open $s\bar{s}$ production from a longitudinally
$(L)$ or transversely $(T)$ polarized photon can be written
as~\cite{adm2,adm4},
\be
\frac{d\sigma^{L(T)}}{dM^2dt}=\frac{2\pi^2\,e_s^2\alpha}{3(Q^2+M^2)^2}
\int\,dz\,\sum_{i,j}\,\mid B^{L(T)}_{ij}\mid^2 ,
\label{cross}
\ee
where $e_s$ is the charge of the strange quark, $\alpha$ is the
fine structure constant,
$M$ is the invariant mass of the $s\bar{s}$ pair, $z\, (1-z)$ is the
light cone fraction of the photon momentum carried by the quark 
(anti-quark) and
$B^{L(T)}_{ij}$ is the helicity amplitude for the 
dissociation of a $L(T)$ polarized photon into a $s\bar{s}$ pair
with helicities $i$ and $j$ respectively. For transversely 
and longitudinally polarized photon the amplitudes 
(for $t=0$) are given by~\cite{adm2},
\ba
{\rm Im}B_{++}^T&=&\frac{m_sI_L}{2h(z)},~~~~
{\rm Im}B_{+-}^T=\frac{-zk_TI_T}{h(z)},\nonumber\\
{\rm Im}B_{-+}^T&=&\frac{(1-z)k_TI_T}{h(z)},~~~~
B_{--}^T=0,\nonumber\\
{\rm Im}B_{+-}^L&=&-{\rm Im}B_{-+}^L=
\sqrt{\frac{Q^2}{2}}h(z)I_L,~~~~B_{++}^L=B_{--}^L=0,
\label{amplitudes}
\ea
where $h(z)=\sqrt{z(1-z)}$ and
\be
I_L=K^2\int^{K^2}\,\frac{dl_t^2}{l_t^4}\alpha_s(l_t^2)f(x,x^\prime,l_t^2)
\left(\frac{1}{K^2}-\frac{1}{K_l^2}\right) ,
\ee
\be
I_T=\frac{K^2}{2}\int^{K^2}\,\frac{dl_t^2}{l_t^4}\alpha_s(l_t^2)
f(x,x^\prime,l_t^2)\left(\frac{1}{K^2} -\frac{1}{2k_T^2}
+\frac{K^2-2k_T^2+l_t^2}{2k_T^2K_l^2}\right) .
\ee
$k_T$ ($-k_T$) is the transverse momentum of
the quark (anti-quark), $K^2=z(1-z)Q^2+k_T^2+m_s^2$, is the scale probed 
by the process and $K_l^2=\sqrt{(K^2+l_t^2)^2-4k_T^2l_t^2}$.
$f(x,x^\prime,l_t^2)$ is the skewed (off-diagonal) gluon  
distribution un-integrated over its transverse momentum, $l_t$.
$x\simeq (Q^2+M^2)/(W^2+Q^2)$ and $x^\prime\simeq\,
(M^2-m_V^2)/(W^2+Q^2)(<<x$),
which indicates that heavier the mesons more important is the
skewness. 
In the present article we use diagonal gluon distribution,
$g(x,l_t^2)$, related to $f(x,l_t^2)$ as follows,
\be
f(x,l_t^2)=\frac{\partial(xg(x,l_t^2))}{\partial \ln l_t^2} .
\ee
The skewness of the gluon distribution has been taken into
account by multiplying the amplitudes by a factor $R_g$~\cite{adm2},
\be
R_g=\frac{2^{2\lda+3}\Gamma(\lda+\frac{5}{2})}{\sqrt{\pi}\Gamma(\lda+4)},
\ee
where $\lda\,\sim\,{\partial\,\ln (xg(x,Q^2))}/{\partial\,\ln (1/x)}$. 

The contribution from the infrared region has been obtained
by introducing an infrared separation scale, $l_0^2$~\cite{adm4}.
\be
I_L=\alpha_s(l_0^2)xg(x,l_0^2)\left(\frac{1}{K^2}-\frac{2k_T^2}{K^4}\right)
+K^2\int_{l_0^2}^{K^2}\,\frac{dl_t^2}{l_t^4}
\alpha_s(l_t^2)f(x,x^\prime,l_t^2)
\left(\frac{1}{K^2}-\frac{1}{K_l^2}\right) ,
\ee
and similarly 
\ba
I_T&=&\alpha_s(l_0^2)xg(x,l_0^2)
\left(\frac{1}{K^2}-\frac{k_T^2}{K^4}\right)
+\frac{K^2}{2}\int_{l_0^2}^{K^2}\,\frac{dl_t^2}{l_t^4}\alpha_s(l_t^2)
f(x,x^\prime,l_t^2)\nonumber\\
&&\times\left(\frac{1}{K^2} -\frac{1}{2k_T^2}
+\frac{K^2-2k_T^2+l_t^2}{2k_T^2K_l^2}\right) .
\ea

The amplitudes, $B_{ij}^{L(T)}$, given above are
evaluated in the proton rest frame. 
As the formation of the $\phi$ takes place in the 
rest frame of the $s\bar{s}$, it is required
to transform the helicity amplitude from the proton rest
frame to the $s\bar{s}$ rest frame through the transformation,
\be
A_{kl}=\sum_{i,j}c_{ik}c_{lj}B_{ij} ,
\ee
where 
\be
c_{++}=c_{--}=c_{+-}=-c_{-+}=\sqrt{(1-a\cdot\,b)/2} ,
\ee
$a_\mu$ is the quark polarization vector in the $s\bar{s}$
rest frame and $b_\mu$ is the corresponding quantities
in the proton rest frame~\cite{adm3}. 

Having obtained these values for the amplitudes in the $s\bar{s}$
rest frame we take the projections of these amplitudes 
in the $J^P=1^-$ states by the following equation,  
\be
A^{L(T)}_{jk}=\sum_{J}\,e_J^{L(T)}d^J_{1\beta} ,
\ee
where $d^J_{1\beta}$ are the spin rotation matrices~\cite{edmond}
and $e_J^{L(T)}$ could be obtained by inverting the above relation.

The amplitudes given in eqs.~(\ref{amplitudes}) contain
only the imaginary part, while the real parts are obtained by using
the relation ${\rm Re}\,A={\rm tan}(\pi\lda/2)\,{\rm Im}A$~\cite{adm3}.
In the present work the NLO correction has been 
taken into account by multiplying the amplitudes
by a $K$ factor, $K=\exp(\pi\,C_F\alpha_s/2)$, where
the scale used as the argument of $\alpha_s$ is $2K^2$ and
$C_F=4/3$.
\section*{III. Results}

The cross section for the $\phi$ production from 
longitudinally and transversely polarized photon
are obtained  by integrating eq.~(\ref{cross})
(with the amplitudes projected in the $J^P=1^{-}$ state)
over an mass interval 1.0 GeV$\,\le\,M\,\le\,$1.04 GeV,
as the $\phi$ has been experimentally observed in this 
invariant mass interval through $\phi\,\ra\,K\,\bar{K}$
decay~\cite{h1phi}. 
The $t$ integration
has been performed by assuming a $t$ dependence of 
the cross section $\sim\,\exp(bt)$, 
with an average slope, $b=5.2$ GeV$^{-2}$, 
taken from experiment~\cite{h1phi}. 

\begin{figure}
\centerline{\psfig{figure=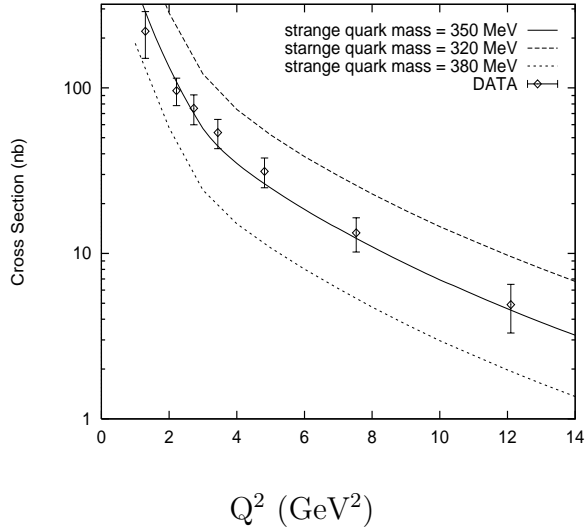,height=6cm,width=8cm}}
\begin{center}
Q$^2$ (GeV$^2$)
\end{center}
\caption{Diffractive $\phi$ production cross section as
a function of $Q^2$ at HERA for GRV98(NLO) 
gluon distribution  and for three values of 
the strange quark masses, $m_s=320,\,350$ and $380$\, MeV. 
The {\it cm} energy of the $\gamma^\star\,p$ system is
 $W=75$ GeV and $l_0^2=1$ GeV$^2$. 
}
\label{xgrv98}
\end{figure}
The typical value of $x$ sampled in diffractive $\phi$
production for $W=75$ GeV is $x\,\sim\,2\times\,10^{-4}$.
For such a low value of $x$ there is a large ambiguity
among various parametrizations of the gluon 
distribution~\cite{cteq,grv98,mrs99}. 
Therefore, we will show the sensitivity
of our results on the gluon distributions.
We start with the GRV98(NLO) gluon distribution.
In fig.~\ref{xgrv98}, the $Q^2$ dependence of the cross section 
is depicted. For the strange quark mass, $m_s=350$ MeV, 
and the infrared scale, $l_0^2=1$ GeV$^2$, the 
agreement between the QCD
based description and the experimental data is reasonably good.
The sensitivity of the 
cross section on the strange quark mass is evident from the figure.
The data can be fitted by appropriately increasing (decreasing)
the invariant mass interval for $m_s=380 (320) $ MeV, but 
in the present work we prefer to fix the window in 
the range $1\,\le\,M\,\le\,1.04\,$ GeV because of the
reason mentioned earlier.
With current quark mass, $m_s\sim\, 150 $ MeV,
it is observed that the theoretical results overestimate 
the data by a large amount with the above invariant mass window.

We note that the experimental data is well reproduced with 
 the strange quark mass, $m_s\,\sim\,350$ MeV which is
 intermediate between the constituent mass ($M_s \sim 500$ MeV)
 and  the current  mass ($m_{s,0} \simeq 150 $ MeV). 
 At this point we recall the momentum-dependent
  effective strange-quark mass $m_s (p)$, 
 which interpolates between the constituent mass and the current mass,
 may be realized through the dynamical breaking
 of chiral symmetry~\cite{hg,hp,kh}.
 In particular, for large space-like momentum
  $p^2 = -P^2 < 0$ ($P^2$: large), the operator product expansion for the
   quark propagator yields the asymptotic behavior \cite{hp}
\be
m_s(P)=m_{s,0}(\mu)\left(\frac{\alpha_s(P)}{\alpha_s(\mu)}\right)^{d}+
\frac{16 \pi \alpha_s (P)}{P^2} | \langle\, \bar{\psi}\psi(\mu)\,\rangle |
\left(\frac{\alpha_s(P)}{\alpha_s(\mu)}\right)^{-d},
\label{mass}
\ee
  where $d (=12/27$ for $N_f=3)$
  is the mass anomalous dimension, $\mu$ is the  
   renormalization point, and 
   $\langle\, \bar{\psi}\psi(\mu)\,\rangle $ is the
    chiral vacuum condensate.
 
\begin{figure}
\centerline{\psfig{figure=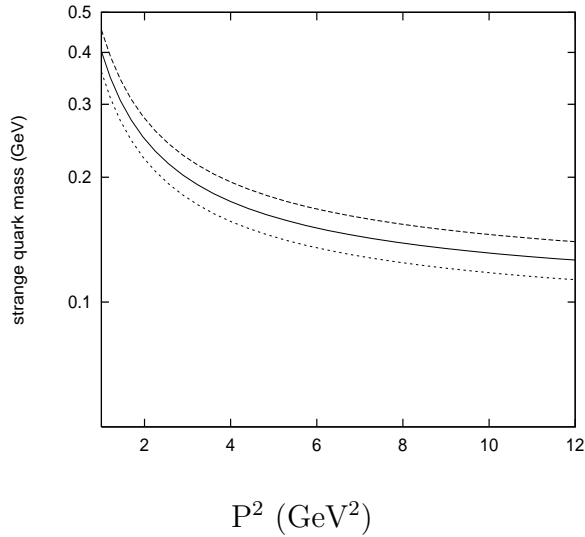,height=6cm,width=8cm}}
\begin{center}
P$^2$ (GeV$^2$)
\end{center}
\caption{The variation of the strange quark mass as a function of $P^2$
 taking into account the uncertainty of the current masses,
  $m_{u,0}$(2GeV)$=3.5\pm 0.4$ MeV, $m_{d,0}$(2GeV)$=6.3\pm 0.8$ MeV and
$m_{s,0}$(2GeV)$=118.9\pm 12.2$ MeV (see text). 
These current  masses are taken from Ref.~\protect\cite{sn}.  
}
\label{rg}
\end{figure}

     In Fig.~\ref{rg}, the asymptotic form of $m_s(P)$ as a function of
 the space-like momentum $P$ is shown, where 
$m_{s,0}$(2GeV)=$118.9\pm 12.2$ MeV,
$m_{u,0}$(2GeV)=$3.5\pm 0.4 $ MeV,  
$m_{d,0}$(2GeV)=$ 6.3\pm 0.8$ and  
 $m_{\pi}^2f_{\pi}^2 \simeq -(m_u+m_d) \langle\,\bar{\psi}\psi\,\rangle $
are taken from the first Ref. of~\cite{sn}.
 To obtain $m_s(P)$ for entire domain of 
  space-like and time-like momenta, 
   one needs to solve the Schwinger-Dyson 
  equation for quark propagator with suitable assumption 
  on the gluon propagator and the quark-gluon vertex at
   low energies (see, e.g., \cite{kh,abra,RW}).  
 In that case, $m_s(P)$ is expected to be a smooth interpolation
  between the asymptotic behavior eq.(\ref{mass}) at large $P^2$ and
   the constituent mass, $M_s \simeq 500$ MeV 
  at $P \sim 0$.
  In our diffractive process, we need  $m_s(P)$
   in the entire domain of $P$ in principle, since 
   the quarks with space-like momentum are initially produced by the
     space-like photon $(Q^2 <0$) and they eventually  
    become time-like after the {\em kick} by the gluons inside the
     proton (see Fig.1). The quark mass, we have found in our 
      analysis, 
    $m_s \sim 350$ MeV, which is smaller
     than $M_s$ but is larger than $m_{s,0}$,
      may be thus interpreted as an effective
     mass averaged over momentum relevant for
      the diffractive process shown in Fig.1. 
   
\begin{figure}
\centerline{\psfig{figure=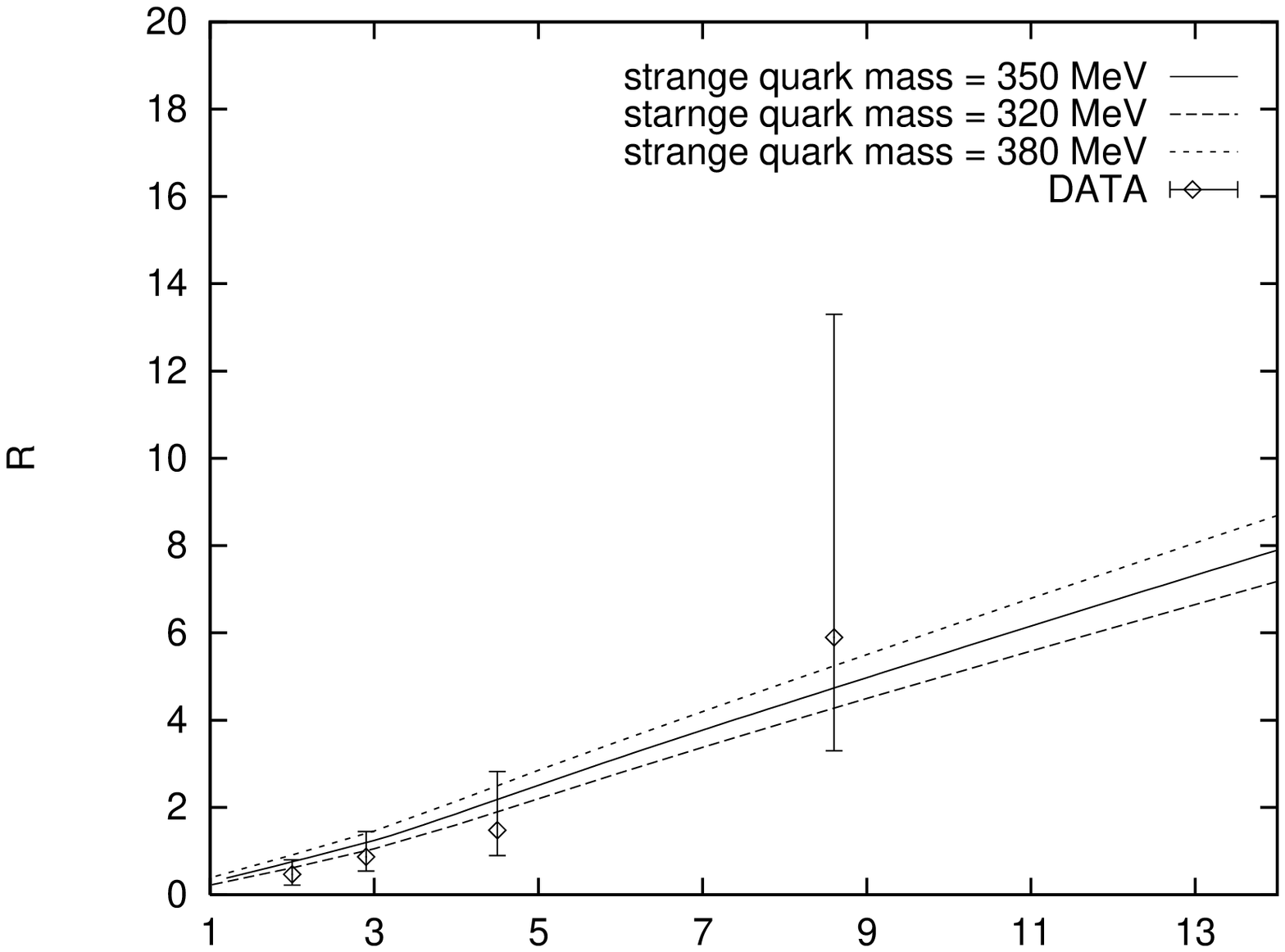,height=6cm,width=8cm}}
\begin{center}
Q$^2$ (GeV$^2$)
\end{center}
\caption{The ratio $R=\sigma_L/\sigma_T$ as a function of $Q^2$
for GRV98(NLO) gluon distribution for $l_0^2=1$ GeV$^{2}$. 
}
\label{rgrv98}
\end{figure}
\begin{figure}
\centerline{\psfig{figure=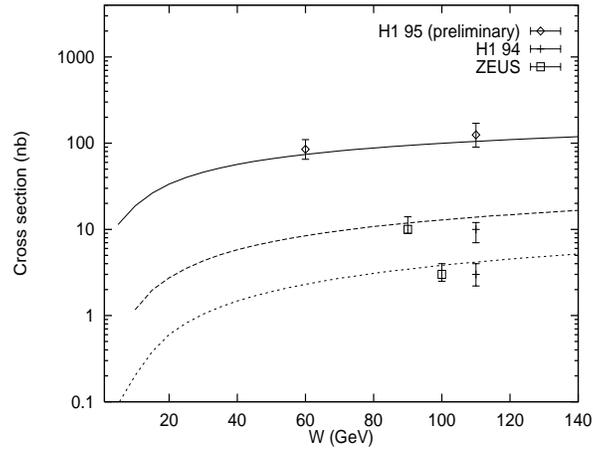,height=6cm,width=8cm}}
\caption{The cross section for $\phi$ production as a function of $W$
for $Q^2=$ 2.5 (solid line), 8.3 (dashed line) and 14.6 GeV$^2$ (dotted line)
for GRV98(NLO) gluon distribution with $l_0^2=1$ GeV$^2$ and $m_s=350$ MeV.
}
\label{epw}
\end{figure}

\begin{figure}
\centerline{\psfig{figure=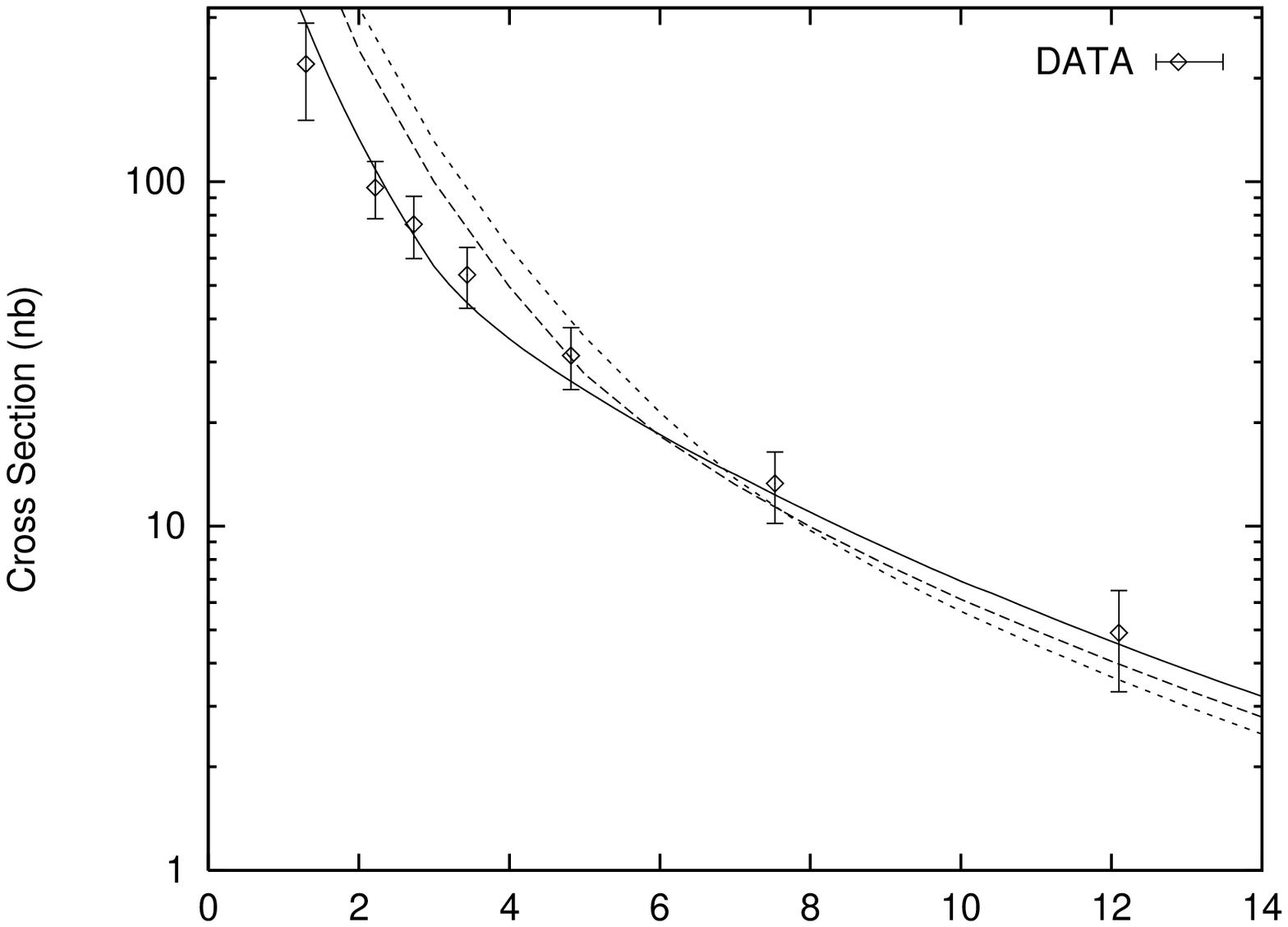,height=6cm,width=8cm}}
\begin{center}
Q$^2$ (GeV$^2$)
\end{center}
\caption{The cross section for $\phi$ production as a function of $Q^2$
for $W=75$ GeV for GRV98 (NLO) gluon distribution for
$l_0^2=$1 (solid line) , 1.5 (dashed line), and 2 
GeV$^2$ (dotted line) with $m_s=350$ MeV. 
}
\label{figl0}
\end{figure}

In fig.\ref{rgrv98} we show the ratio $R=\sigma_L/\sigma_T$
as a function of $Q^2$. The theoretical calculation shows
the correct trend.
Putting the constraint on the strange quark mass and the infrared scale 
from the experimental data shown in fig.\ref{xgrv98}, we evaluate
the  $W$ dependence of the cross section for various values of $Q^2$.
The agreement between the experimental data (taken from~\cite{ha})
and the theoretical calculation 
is satisfactory. 
In fig.\ref{figl0} we show the results for 
various values of $l_0^2$ with $m_s=350$ MeV. 
Results obtained for $l_0^2=1.5$ and $2$
GeV$^2$ start deviating from the experimental value at small $Q^2$.

\begin{figure}
\centerline{\psfig{figure=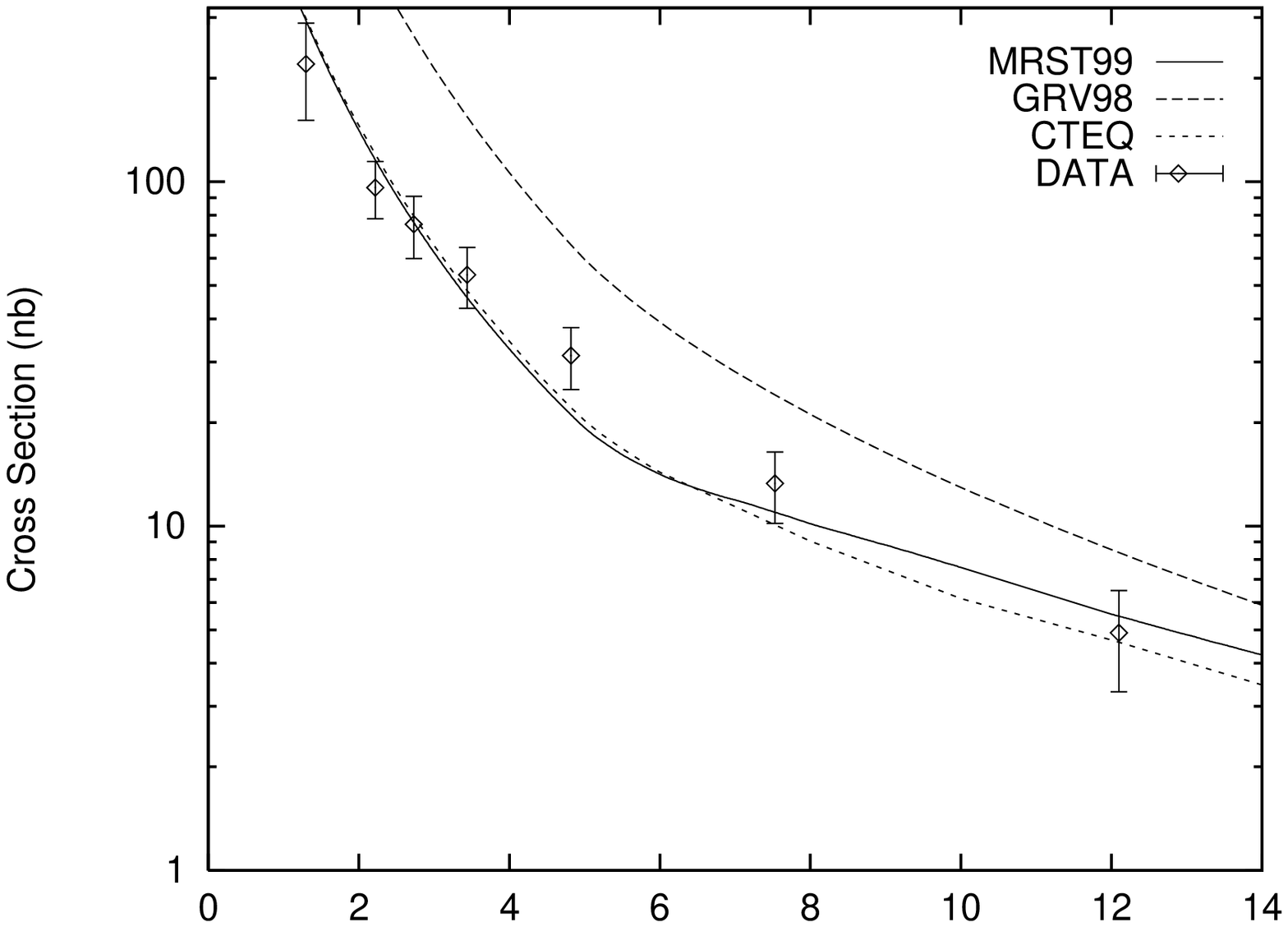,height=6cm,width=8cm}}
\begin{center}
Q$^2$ (GeV$^2$)
\end{center}
\caption{Diffractive production of $\phi$ at HERA for various
parametrization of gluon distribution function. 
 $m_s=320$ MeV and  $l_0^2=1.5$ GeV$^2$ are adopted. 
}
\label{xgluons}
\end{figure}
In fig.\ref{xgluons} we show the $Q^2$ behavior of the
cross section for different gluon distribution functions.
The value of the infrared scale, $l_0^2$ and 
strange quark mass, $m_s$ are $1.5$ GeV$^2$ and 320 MeV
respectively. Although GRV98(NLO) gluon distribution overestimates,
the predictions with MRST99 and CTEQ5M gluon distributions
 are in agreement 
with the experimental results.
\begin{figure}
\centerline{\psfig{figure=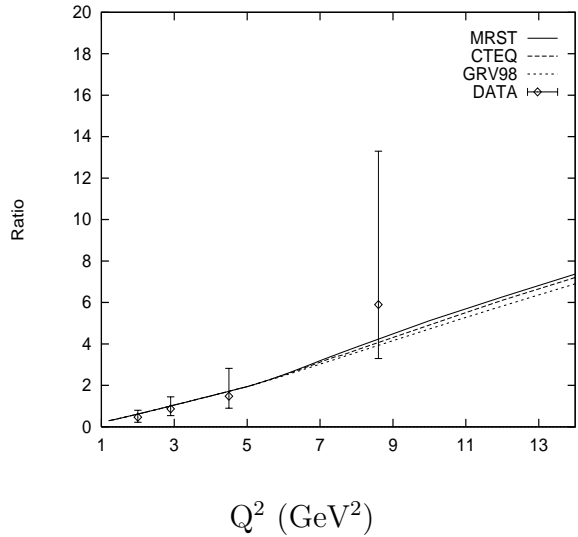,height=6cm,width=8cm}}
\begin{center}
Q$^2$ (GeV$^2$)
\end{center}
\caption{The ratio $R=\sigma_L/\sigma_T$ as a function of $Q^2$
for various parametrization of the gluon distribution function with
$m_s=320$ MeV and  $l_0^2=1.5$ GeV$^2$. 
}
\label{rgluons}
\end{figure}
With smaller values of $m_s\,\sim\,150$ MeV, however, we fail 
to describe the data provided the invariant mass window
is kept fixed at $1\,\le\,M\,\le\,1.04$ GeV, where
the $\phi$ has been measured experimentally.
Fig.~\ref{rgluons} indicates the variation of $\sigma_L/\sigma_T$
as function of $Q^2$ for CTEQ5M, GRV98(NLO) and MRST gluon distributions,
 the ratio
seems to be less sensitive to the gluon distributions.
\vskip 0.1 in
\section*{IV. Summary and Conclusions}
We have studied the diffractive $\phi$ production 
at HERA energies measured by H1 collaboration 
within the ambit of pQCD model based on parton 
hadron duality. The effects of the off-diagonal gluon
distribution and the NLO corrections through the
$K$-factor have been incorporated.
The sensitivity of the results
on the infrared separation scale and
the various parametrization of the gluon distribution
have been discussed.
It is found that, with reasonable choice
 of the infrared scale, the total
cross section and the ratio, $\sigma_L/\sigma_T$
are well described in the present framework with
 an effective strange quark mass, $m_s\,\sim\,320-380$ MeV. 
 Such a value of the strange quark mass, which lies
between constituent and  current quark masses may be closely
 related to the momentum-dependent dynamical mass
  associated with the dynamical breaking of chiral symmetry in QCD.
 
\vspace{1cm}

\noindent{{\bf Acknowledgement}: 
We are grateful to K. Itakura for useful discussions. 
K.S. and A. H. are supported by JSPS Research Fellowship for
Young Scientists. J.A. is grateful to the Japan
Society for Promotion of Science (JSPS) for financial support.
J.A. and T.H. are also supported by Grant-in-aid for Scientific
Research No. 98360 of JSPS.

\end{document}